\documentclass[
    ,final            % use final for the camera ready runs
  ]
  {aipproc}

\layoutstyle{8x11single}

\usepackage{amssymb,epstopdf,amsmath}

\begin{document}

\def\nuc#1#2{${}^{#1}$#2}

\title{A Radon Progeny Deposition Model}


\author{V. E. Guiseppe}{
  address={University of South Dakota, Vermillion, South Dakota 57069}
}

\author{S. R. Elliott}{
  address={Los Alamos National Laboratory, Los Alamos, New Mexico 87545}
}

\author{A. Hime}{
  address={Los Alamos National Laboratory, Los Alamos, New Mexico 87545}
}

\author{K. Rielage}{
  address={Los Alamos National Laboratory, Los Alamos, New Mexico 87545}
}

\author{S. Westerdale}{
  address={Los Alamos National Laboratory, Los Alamos, New Mexico 87545}
  ,altaddress={Massachusetts Institute of Technology, Cambridge, Massachusetts 02139} % additional visiting address
}

\begin{abstract}
The next generation low-background detectors operating underground aim for unprecedented low levels of radioactive backgrounds. Although the radioactive decays of airborne radon (particularly \nuc{222}{Rn}) and its subsequent progeny present in an experiment are potential backgrounds, also problematic is the deposition of radon progeny on detector materials. Exposure to radon at any stage of assembly of an experiment can result in surface contamination by progeny supported by the long half life (22 y) of \nuc{210}{Pb} on sensitive locations of a detector. An understanding of the potential surface contamination from deposition will enable requirements of radon-reduced air and clean room environments for the assembly of low background experiments. It is known that there are a number of environmental factors that govern the deposition of progeny onto surfaces. However, existing models have not explored the impact of some environmental factors important for low background experiments. A test stand has been constructed to deposit radon progeny on various surfaces under a controlled environment in order to develop a deposition model. Results from this test stand and the resulting deposition model are presented.

\end{abstract}

\maketitle

%%%%%%%%%%%%%%%%%%%%%%%%%%%%%%%%%%%%%%%%%%%%
%% MAINMATTER
%%%%%%%%%%%%%%%%%%%%%%%%%%%%%%%%%%%%%%%%%%%%

\section{Introduction}

\begin{figure}[h]
\includegraphics[width=0.65\textwidth]{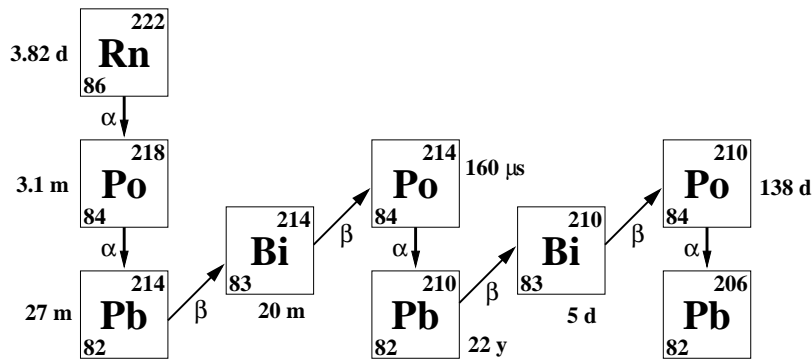}
\caption{The \nuc{222}{Rn} decay chain. The rare branches to \nuc{218}{At} and \nuc{210}{Tl} are removed for clarity.}
\label{fig:chain}
\end{figure}

Observation of rare processes demands a careful study and implementation of signal processing and detector design to reach new levels of background suppression. The potential backgrounds present must be thoroughly understood so that backgrounds as rare as the signal of interest can be predicted and modeled. One problematic class of backgrounds are from radon gas. 
Radon is a naturally occurring radioactive gas present both underground and above ground. Radon gas cannot be tolerated underground in the laboratory of an ultra-low background experiment. Both radon and some of its short-lived progeny (Fig. \ref{fig:chain}) emit a host of background radiation and nuclear recoils  that can mask a desired rare event signal. Radon gas must be mitigated for ultra-low background experiments in underground laboratories. In addition, exposure to radon at any stage of assembly of an experiment can result in surface contamination by progeny supported by the long half-life (22 y) of \nuc{210}{Pb} on sensitive surfaces of a detector \cite{ams07, leu05}. The radon progeny surface contaminants can continue to produce unwanted background even after the detector is moved to its final laboratory with the low radon level. For the case of a dark matter detector, the $\alpha$ decay of \nuc{210}{Po} produces a nuclear recoil similar to that of a WIMP-nucleon scatter. This is a particularly problematic background as it closely mimics the expected signal. The energetic  $\alpha$ and $\beta$ decays of \nuc{210}{Po} and \nuc{210}{Bi} respectively produce a background near or above the region of interest for neutrinoless double-beta decay searches and can also mimic the expected signal.

% cut?
%For the case of a dark matter detection, the $\alpha$ decay of \nuc{210}{Po}  produces a nuclear recoil similar to that of a WIMP-nucleon scatter. This is a particularly problematic background as it closely mimics the expected signal. A neutrinoless double-beta decay experiment can be susceptiple to the emitted ionization for the 

Few models exist in the literature to describe radon progeny deposition. One  model suitable for indoor air \cite{naz98} states that the rate of deposition ($R_d$) is a function of the radon concentration ($C$), the surface area ($S$), and the deposition velocity ($v_d$)
\begin{equation}
R_d = v_d S C
\label{eq:naz}
\end{equation}
The  deposition velocity factor includes the mechanisms of progeny deposition onto a surface and is found to vary by orders of magnitude due to the environmental conditions of the air. Though this model has been found to explain the deposition of radon progeny in indoor air, it is not sufficient to make predictions of radon progeny deposition relevant to low background experiments. The critical assembly of  an experiment is commonly performed in a clean room and/or a glove box, which has an environment far from that found in typical indoor room air. Instead, there is often frequent air circulation and exchanges with active HEPA filtration to meet a desired cleanliness. In addition, the deposition velocity does not differentiate how the mechanism of deposition relates among various materials.

The demand to study radon progeny deposition in a clean room environment was recognized by many low background experiments including the Borexino Collaboration \cite{ben07}. A study was performed to quantify and directly measure the deposition velocity of progeny on nylon in a mock clean room environment \cite{leu05}. They found the deposition velocity to be many orders of magnitude lower than that found by Ref. \cite{naz98}. The lower deposition rate is explained to be caused by the clean room air circulation and filtration. While these results were specific for their individual needs, a more general model would be useful to other low background experiments.

\section{Radon deposition measurements}

\begin{figure}
\includegraphics[width=0.4\textwidth]{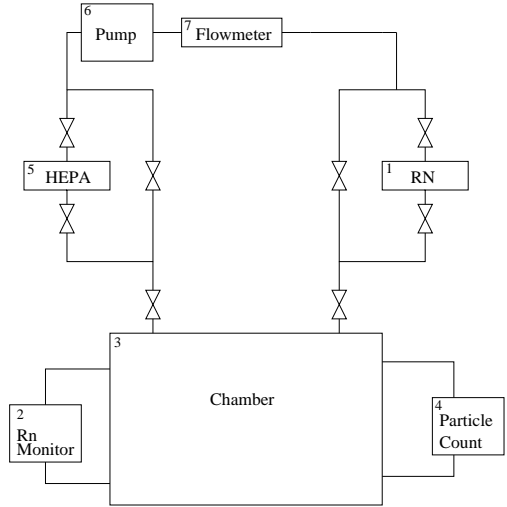}
\caption{A schematic of the current radon progeny deposition test stand showing a 1) Rn source, 2) Rn monitor, 3) exposure chamber, 4) particle counter, 5) HEPA filter, 6) circulating pump and control valves.}
\label{fig:radons}
\end{figure}

We have assembled a test stand to directly measure the deposition rate of radon progeny under various environmental conditions for several materials. The test stand allows the development of models describing the factors controlling the deposition of radon progeny on surfaces of materials. In order to develop a multi-dimensional deposition model, we intend to look at significant sources of variability in the deposition rate including: exposure time, suspended particle concentration, radon concentration, HEPA filtering and flow/exchange rate, materials and surface preparation, electrostatics, temperature, humidity, etc. The test stand is made from a closed system where radon gas from a Pylon flow-through source is introduced into a exposure chamber along with HEPA filtration (99.97\% at 0.3 $\mu$m), radon monitoring (Durridge Co. RAD7), and particle count monitoring shown schematically in Fig. \ref{fig:radons}.

In order to directly measure the deposition rates of radon progeny, material samples are exposed to radon to allow a build-up of a surface contamination. The number ($N_i$) of the progeny present on a surface at time $t$ is governed by deposition and decay terms and is given by

\begin{align}
 \frac{dN_1}{dt} &= d_1 S C - \lambda_1 N_1 \label{eq:depostart}\\
\frac{dN_2}{dt} &= d_2 S C + (1 - r) \lambda_1 N_1 - \lambda_2 N_2 \\
\frac{dN_3}{dt} &=  d_3 S C + \lambda_2 N_2 - \lambda_3 N_3 \label{eq:depostop}
\end{align}
where $d_i$ is the atomic deposition rate, $N_i$ is the number of atoms, and $\lambda_i$ is the decay rate of the nuclides  \nuc{218}{Po}, \nuc{214}{Pb}, and \nuc{214}{Bi}. With the radon concentration ($C$) in units of [Bq m$^{-3}$] and the surface area ($S$) in [m$^2$], the deposition rate in [m min$^{-1}$ Bq$^{-1}$] is quantitatively the same as the deposition velocity in Equation (\ref{eq:naz}). The fraction of \nuc{218}{Po} decays in which the recoil of the atom ejects itself from the surface is given by $r$.  The recoils from the beta decays of Pb and Bi are not energetic enough to eject the nucleus from the surface. The longest half-life of the first four radon progeny is 27 min. and therefore equilibrium between radon and its progeny is achieved within a few hours. The deposition rate of  \nuc{214}{Po} is assumed to be negligible due to its short half-life of 160 $\mu$s. 

At the conclusion of an exposure, the material samples are removed from the high radon environment thereby removing the deposition term. The decay rates after exposure are given by
\begin{align}
\frac{dN_1}{dt} &= - \lambda_1 N_1 \label{eq:decaystart}\\
\frac{dN_2}{dt} &= (1 - r) \lambda_1 N_1 - \lambda_2 N_2 \\
\frac{dN_3}{dt} &=  \lambda_2 N_2 - \lambda_3 N_3 \\
\frac{dN_4}{dt} &=  \lambda_3 N_3 - \lambda_4 N_4 \label{eq:decaystop}
\end{align}
where $N_4$ is the number of \nuc{214}{Po} atoms. 

\begin{figure}
\includegraphics[width=0.55\textwidth]{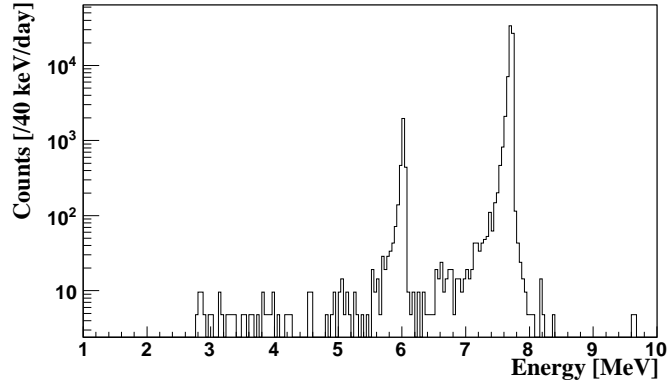}
\caption{An example $\alpha$ decay energy spectrum from \nuc{218}{Po} at 6.0 MeV and \nuc{214}{Po} at 7.8 MeV.}
\label{fig:alpha}
\end{figure}
Materials are exposed to varying radon concentrations with fixed environmental conditions. The exposed materials discussed here include acrylic and Cu disks with an exposed surface area of 20 cm$^2$. Exposures were conducted with a constant circulation flow rate and the HEPA filter arrangement in the system allows for filtered or unfiltered operation. The samples are exposed to a known radon concentration for a fixed amount of time followed by an immediate $\alpha$ spectroscopy measurement to count the progeny present after exposure. An $\alpha$ counter (Ortec 1200 mm$^2$ ULTRA AS silicon detector)  measures the two deposited, short-lived $\alpha$ emitters in the Rn chain: \nuc{218}{Po} and \nuc{214}{Po}. The two Po $\alpha$ peaks are separated by 1.8 MeV and are clearly resolved by the $\alpha$ spectrometer (Fig. \ref{fig:alpha}) with a nominal FWHM energy resolution of 40 keV. The low energy tails in the figure show the result of  $\alpha$ straggling prior to charge collection in the detector. A peak sensing ADC provides the pulse height of the shaped $\alpha$ counter signal and a timestamp for offline analysis. The counting efficiency of the $\alpha$ counter is determined by a Monte Carlo simulation.

\subsection{Results of Deposition}

\begin{figure}
\includegraphics[width=0.95\textwidth]{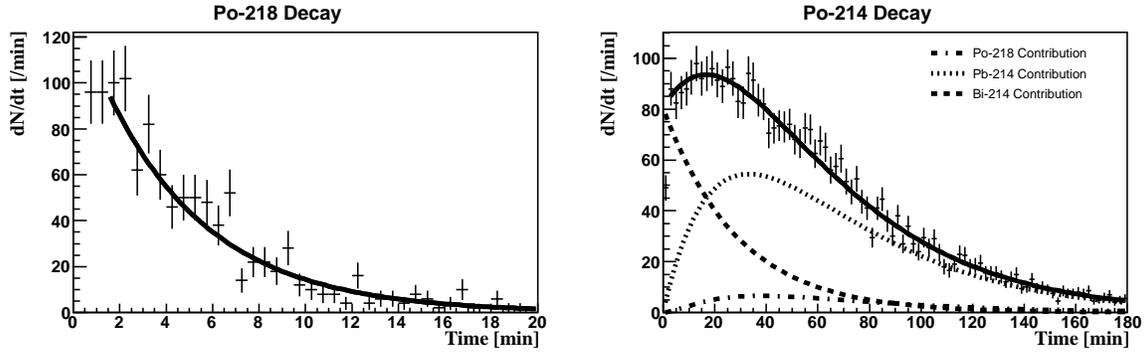}
\caption{The $\alpha$ decay curves for unsupported decay of \nuc{218}{Po} and the supported decay of \nuc{214}{Po} after a sample of acrylic was exposed to 220 kBq/m$^3$ of radon for 60 min. The \nuc{214}{Po} decay includes contributions from initial concentration of \nuc{218}{Po}, \nuc{214}{Pb}, and \nuc{214}{Bi}.}
\label{fig:decay}
\end{figure}

\begin{figure}
\includegraphics[width=0.55\textwidth]{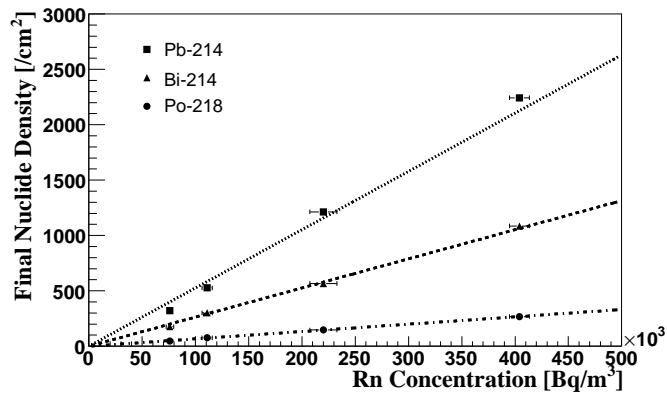}
\caption{The final progeny density remaining on exposed acrylic under an unfiltered flow rate of 2.5 L/min as a function of radon concentration. The slopes provide the three progeny deposition rates for the stated conditions.}
\label{fig:fit}
\end{figure}

The solutions to Equations (\ref{eq:decaystart}-\ref{eq:decaystop}) fit to the measured data provides an estimate of the number of progeny at the end of exposure. At present, we assume $r=0.5$ for the recoil fraction. Examples of decay curves are shown in Fig. \ref{fig:decay} for an exposure of acrylic to radon. 
The solutions to Equations (\ref{eq:depostart}-\ref{eq:depostop}) are fit to the final progeny density remaining on a material at the end of exposure as a function of radon concentration found from the decay curves. This fit provides the deposition rate for the first three progeny and an example  is shown in Fig. \ref{fig:fit} for the case of exposed acrylic under a unfiltered circulation rate of 2.5 L/min. 

\begin{figure}
\includegraphics[width=0.55\textwidth]{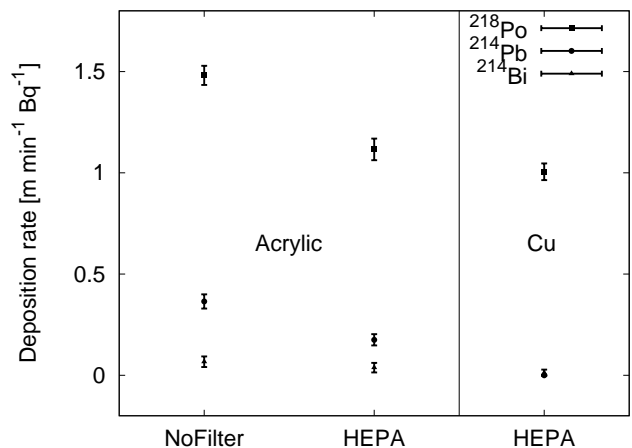}
\caption{The preliminary deposition rate results of \nuc{218}{Po}, \nuc{214}{Pb}, and \nuc{214}{Bi} under unfiltered and HEPA filtered conditions on acrylic and under HEPA filtered conditions on Cu.}
\label{fig:results}
\end{figure}

Preliminary results of the deposition rate of the first three radon progeny are shown in Fig. \ref{fig:results} under unfiltered and HEPA filtered conditions on acrylic and under HEPA filtered conditions on Cu. These data suggest that the HEPA filtration suppresses the deposition of the radon progeny and that little difference in progeny deposition behavior between acrylic and Cu exists. By lowering the progeny concentration in the air, the HEPA filtration effectively reduces the deposition rate as suggested earlier. This initial work underscores the importance of understanding the various factors that control the deposition rate of radon progeny in a radon environment. 

\begin{figure}
\includegraphics[width=0.55\textwidth]{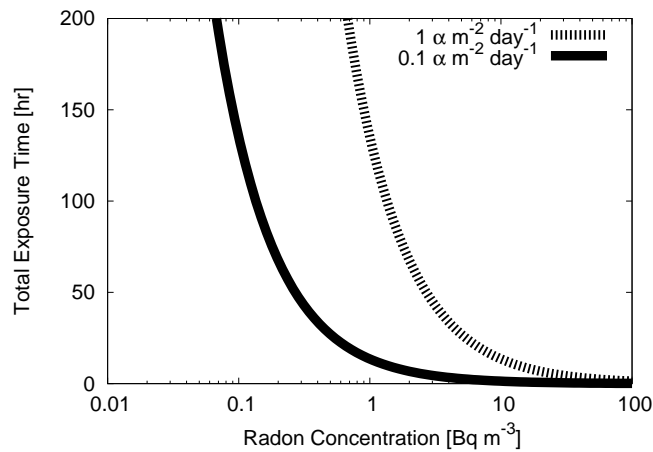}
\caption{An example radon exposure requirement to achieve a desired surface activity of 0.1 and 1.0 $\alpha$ decay m$^{-2}$ day$^{-1}$ due to \nuc{210}{Po} on acrylic in an HEPA-filtered environment.}
\label{fig:req}
\end{figure}

Since the deposition rate of radon progeny on a material is assumed and shown here to be linearly proportional to the exposed radon concentration, projections can be made. For a specified surface activity of a low background experiment, an exposure limit can be determined from measured deposition rates. For example, Fig. \ref{fig:req} shows the exposure limits as a function of the radon concentration for acrylic in a filtered environment for two desired surface activities of \nuc{210}{Po} when the only source of surface contamination is due to the deposition of the first three radon progeny. 

\section{Summary}

A program to directly measure the deposition rate of radon progeny on materials is underway in order to develop a general model of deposition. The preliminary data indicates that the effect of filtration in an  environment has an effect on the radon progeny deposition rate. Future tests will further explore the type of material deposited on and to evaluate the role of electrostatics and electrochemical attractions.

It is important to understand all possible background events from radon and the total effect of exposure of ultra-pure materials to radon gas. Knowledge of  the nuclide  deposition rate onto a particular material under a known set of conditions permits the specification of radon exposure limits to keep the resulting background below a desired level. In addition, optimization of clean room or glove box filtering and flow rates is possible with a thorough understanding of the deposition behavior.

%%%%%%%%%%%%%%%%%%%%%%%%%%%%%%%%%%%%%%%%%%%%%%%%
%% BACKMATTER
%%%%%%%%%%%%%%%%%%%%%%%%%%%%%%%%%%%%%%%%%%%%%%%%

\begin{theacknowledgments}
We thank our DEAP/CLEAN and {\sc Majorana} collaborators for fruitful discussions. This work was made possible by support from the Laboratory Directed Research and Development Program at Los Alamos and the Office of Research at the University of South Dakota.
  
\end{theacknowledgments}

%%%%%%%%%%%%%%%%%%%%%%%%%%%%%%%%%%%%%%%%%%%%%%%%
%% The bibliography can be prepared using the BibTeX program or
%% manually.
%%
%% The code below assumes that BibTeX is used.  If the bibliography is
%% produced without BibTeX comment out the following lines and see the
%% aipguide.pdf for further information.
%%
%% For your convenience a manually coded example is appended
%% after the \end{document}
%%%%%%%%%%%%%%%%%%%%%%%%%%%%%%%%%%%%%%%%%%%%%%%%

%%%%%%%%%%%%%%%%%%%%%%%%%%%%%%%%%%%%%%%%%%%%%%%%
%% You may have to change the BibTeX style below, depending on your
%% setup or preferences.
%%
%%
%% For The AIP proceedings layouts use either
%%%%%%%%%%%%%%%%%%%%%%%%%%%%%%%%%%%%%%%%%%%%

\bibliographystyle{aipproc}   % if natbib is available
%\bibliographystyle{aipprocl} % if natbib is missing

%%%%%%%%%%%%%%%%%%%%%%%%%%%%%%%%%%%%%%%%%%%
%% You probably want to use your own bibtex database here
%%%%%%%%%%%%%%%%%%%%%%%%%%%%%%%%%%%%%%%%%%%
\bibliography{/Users/vguiseppe//work/latex/bib/myradon}

\end{document}